\documentclass[aps,pra,preprint,showpacs]{revtex4}
\usepackage{amsmath,graphicx,epsfig}
\begin{document}
\def\la{{\langle}}
\def\ra{{\rangle}}

\title{Weak values, 'negative probability' and the uncertainty
principle.}
%
%
\author{D.Sokolovski}
\address{School of Mathematics and Physics,
  Queen's University of Belfast,
  Belfast, BT7 1NN, United Kingdom,}

   \date{\today}
   \begin{abstract}

A quantum transition can be seen as a result
of interference between various pathways
(e.g. Feynman paths) which can be labelled 
by a variable $f$.
An attempt to determine the value of  $f$ without
destroying the coherence between the pathways
produces the weak value of $\bar{f}$.
We show $\bar{f}$ to be an average obtained
with amplitude
distribution which can, in general, take
negative values which, in accordance with
the uncertainty principle, need not contain
information about the actual range of $f$ 
which contribute to the transition.
It is also demonstrated that the moments
of such alternating distributions
have a number of unusual 
properties which may lead to
misinterpretation of the weak
measurement results.
We provide a detailed analysis of weak 
measurements with and without post-selection.
Examples include the double
slit diffraction experiment,
 weak von Neumann and von Neumann-like
measurements, traversal time for an elastic
collision, the phase time, the local angular momentum
(LAM)  and  the 'three-box case' of {\it Aharonov et al}.

\end{abstract}

%
%
\pacs{PACS number(s): 03.65.Ta, 73.40.Gk}
\maketitle

\section{Introduction}
 In his book with Hibbs \cite{Feyn}
 Feynman formulates the uncertainty principle
 as follows:  'Any determination of an alternative
 taken by a process capable of following more
 than one a alternative destroys the inteference
 between alternatives'.
Thus the converse is true:
for a quantum system which can reach its final
state via a number interfering pathways 
the uncertainty principle forbids specifying
which of the routes has actually been taken.
 The latter can be quantified as follows:
 if the interfering pathways are labelled
by some variable $f$, the value of $f$ must remain, 
in some sense, indeterminate.
\newline
The concepts of interfering pathways provides 
a convenient description of a general quantum 
measurement \cite{SR1}, \cite{SR2}. For a quantum system prepared in a state $\Psi_0$ and
later observed (post-selected) in a state $\Psi_1$.
the transition amplitude between the states can be written
as a sum over virtual paths traced by some variable
$\hat{A}$ (e.g., Feynman paths, if $A$ represents the coordinate), which can be arranged according to 
the value $f$ of some functional $F[path]$, e.g., the value
of $\hat{A}$ at some intermediate time, or its time average.
The classes form form a discreet or continuous set of pathways
connecting $\Psi_0$ and  $\Psi_1$ and
 the value of $f$ can be determined if a system is subjected to a measurement
which converts,
in full analogy with the double-slit interference experiment \cite{Feyn},
 interfering pathways into exclusive ones.
\newline
One wonders then what would be the 
result of trying to obtain some information
about $f$ while keeping the interference intact.
A straightforward attempt to write down even an average 
answer fails since no probabilities can be ascribed
to the pathways,
while an 'average' formally constructed with the probability amplitudes
$\Phi(f)$
\begin{equation} \label{0.1}
\bar{f} \equiv \sum f \Phi(f)/\sum \Phi (f)
\end{equation}
is complex valued and its physical significance
is not immediately clear.
Alternatively, one can consider a von Neumann measurement with
the interaction between the measured
system and the meter deliberately reduced in order to minimise
the perturbation incurred and analyse the meter's readings.
This general method based on analysing inaccurate,  or weak,
quantum measurements, was originally formulated by Aharonov et al \cite{Ah1,Ah2,AhBOOK}
and is closely related to the attempts to define the duration of a scattering event
still debated in literature \cite{Rev1,Rev2,Rev3}. 
The original approach to the problem, which leads to the so-called 
Wigner-Eisenbad time delay, or the phase time, relies on following the centre of mass
of the scattered wavepacket, \cite{WE,Smith, Rev1, Rev2} and
has been shown to be equivalent to a weak measurement experiment \cite{SMS}.
A different method was proposed 
by Baz'  \cite{Baz1, Baz2, Baz3,BazBOOK} who considered a particle
weakly coupled to a Larmor clock.  Baz' criticised
the phase time of Smith \cite{Smith} as generally incorrect
and pointed out that elastic collision time must be a 
sharply defined quantity \cite{Baz2, BazBOOK}. 
The Larmor time was shown to be a weak value of the 
traversal time functional \cite{SB}
 in \cite {SC1,ST, IAN}.
 and its relation to the complex time obtained with 
 the help of Eq.(\ref{0.1}) was established in \cite{SB,SBOOK}.
 The introduction of such a complex time has often been 
 criticised on the grounds that any observable physical
 quantity must be real (see, for example, \cite{Rev1,Rev2}).
 \newline
 Finally, the local angular momentum (LAM), a quantity 
given by an expression broadly similar to the phase time,
has recently been employed in \cite{LAM} in order
to identify angular momenta  which contribute to an angular
distribution at a given scattering angle.
\newline
 All above examples have a common purpose of determining,
 in some sense,
 the value of a physical variable without choosing between the 
 alternatives which contribute to a transition. The 
 purpose of the present paper is to analyse the origin, properties
 and general usefulness of weak values which occur 
 in various contexts.
 The term 'negative probability'  was introduced by Feynman \cite{FeynN}
 and was used in connection with Wigner functions \cite{Scully, Muck}
 and the tunnelling time \cite{SC3}.
 In a slightly different context of this paper we will show that 
 the weak values are related to 
 alternating 'improper' distributions which arise because
 a probability amplitude may take negative values.
 In agreement with the uncertainty principle, 
 averaging with such distributions
 effectively 'hides' the information about interfering pathways. 
 Relative complexity of such analysis owes to fact
 that while, in the absence of probabilities, certain
 quantities may exhibit obviously wrong values,
 they do not have always  to do.
 Thus, there is  danger of extrapolating between different
 cases each of which needs instead to be analysed separately.
 \newline
 The rest of the paper is organised as follows.
 In Section 2 we list some elementary properties
 of 'improper' non-positive distributions and their 
 moments. 
 Im Sect 3 we consider a inaccurate classical meter 
 and show that
  it is still possible to extract the results of an accurate measurement
  from the meter's readings.
  In Section 4 we consider the quantum version of the classical 
  meter.
  In Sect. 5 we show that, unlike in the classical case,
  results of an accurate quantum measurements cannot 
  be extracted from the readings of a quantum meter
  with a large (quantum) uncertainty in its initial position.
  Rather the results are expressed in terms of the
  mean $\bar{f}$ and the higher moments of the improper
   distribution $\Phi(f)$ in Eq.(\ref{0.1}). 
  In Sect.6 we consider, as an example, the double-slit
  experiment and equivalent measurement on a two-level 
  system.
  In Sect.7 we show that if the reading of an
  inaccurate meter are average over the final states
  of the system, the results are expressed in terms of the 
 mean and variance of  obtained with a real
 distribution  $w_1$ which is not, in general, non-negative.
 In Section 8 we show the impulsive von Neumann
 measurement without post-selection to  be a special 
 case where the distribution $w_1$ is non-negative and
 coincides, as in the classical case of Sect.3,
  with the probability distribution of 
 an accurate measurement.
 In Section 9 we use the results of Sect.7 to show Baz' conclusion that the 
 elastic collision time is sharply defined \cite{Baz2,BazBOOK} to be wrong.
 In Sect.10 we show the Wigner-Esenbad phase time to 
 be a weak value similar to  (\ref{0.1}) and briefly discuss some
 of its anomalous properties.
 In Sect.11 we show the LAM \cite{LAM} to  be a particular kind of a weak value.
 In Sect.12 we briefly discuss the 'three box case' considered in \cite{AhBOOK,3B1}
 and further discussed in \cite{3B2,3B3}.
 We will show  that the  Aharonov, Leibowitz and Bergmann (ABL) rule
 to be a simple consequence of Feynman's rule for ascribing probabilities
 and suggest an alternative interpretation of the 'three box paradox' based on the uncertainty principle.
 Section 13 contains our conclusions.

\section {Proper, improper and complex distributions}
Consider a real function $\rho(f)$, 
contained within the
interval $0\le x \le 1$,
which can be used to construct a normalised distribution,
\begin{equation} \label{1.1}
w(f)\equiv \rho(f)/\int_0^1 \rho(f)df.
\end{equation}
If $\rho(f)$ does not change sign, $w(f)$ is non-negative and 
can, therefore, be used as a proper probability distribution
to calculate various moments of the random variable $f$.
In this Section we will assume that $\rho$  may
change sign within the interval
and list the consequences 
for the moments and averages calculated with such an improper
distribution.
\newline
a) While for $w(f)\ge 0$ the expectation value
\begin{equation} \label{1.2}
\la f \ra \equiv \int_0^1 f \rho(f) df/\int_0^1 \rho(f) df
\end{equation}
always lies in the region containing the support of 
$\rho$, i.e., between $0$ and $1$, for an alteranting $\rho(f)$ 
this is no longer true. as the normalisation integral 
in Eq.(\ref{1.1}) can take either sign or vanish.
For a simple example consider a function
\begin{equation} \label{1.2a}
\rho^{\epsilon}(f) \equiv \sin(2\pi f)+\epsilon\quad for \quad 0\le f \le 1,
\quad and \quad 0 \quad otherwise
\end{equation}
with the first three moments
\begin{equation} \label{1.3}
\int \rho (f)df =\epsilon, \quad \la f \ra =1/2 - 1/2\pi\epsilon, \quad \la f^2 \ra =1/3 - 1/2\pi\epsilon
\end{equation}
which yields an improper distribution for $|\epsilon| \le 1$.
For  $\epsilon \rightarrow \pm 0$, the value of $|\la f 
\ra| \rightarrow
\infty\ra$
becomes anomalously large. 
In general, an improper expectation value no longer gives an
estimate for the location of the support of the corresponding
distribution $w(f)$.
\newline
b) While, for a proper distribution $w(f) \ge 0$, the equality
\begin{equation} \label{1.4}
\la f^2 \ra - \la f \ra^2=0
\end{equation} 
forces the conclusion that $f$ is a 
sharply defined quantity,
\begin{equation} \label{1.5}
w(f) = \delta (f-\la f \ra),
\end{equation}
the variance (\ref{1.4}) may vanish
for a broad alternating 
distribution.
For example, for the $\rho^\epsilon (f)$
in Eq.(\ref{1.2a}) this would be the case for
\begin{equation} \label{1.7}
\epsilon = \pm 3^{1/2}/\pi
\end{equation}
In general, examining the first two moments of a (possibly) improper 
distribution does allow to establish whether the variable
$f$ is sharply defined.
\newline
c)  for a proper probability distribution the fact that
\begin{equation} \label{1.8}
\int_a^b w(f)df =0,
\end{equation}
where $[a,b]$ lies inside the interval $[0,1]$, guarantees that 
the support of $\rho(x)$ is contained between $0$ and $a$ and $b$ and $1$,
while for an improper distribution making such an assumption leads to an
obvious contradiction.
For example, for
\begin{equation} \label{1.9}
w(f)=3\pi/2 sin(3\pi f)
\end{equation}
we have $\int_0^{2/3}w(f)df =\int_{1/3}^{1}w(f)df=0$
and adopting the above reasoning we must conclude that
$\rho(f)$ with certainty takes its values in each of two
different regions, $0\le f \le 2/3$ and $1/3 \le f \le 1$.
Moreover , there is an infinite number of ways 
to construct subintervals of $[0,1]$ in each of which
the integral (\ref{1.8}) would vanish.
Thus, we cannot, in general, uniquely determine
which part of the interval $0 \le f \le 1$ contributes to
the normalisation integral $\int_0^1 \rho(f) df$
even though parts of the interval must be redundant
due to the cancellation,
and have to conclude that all values of $f$ between 
$0$ and $1$ are equally important. Relation between 
this  observation and the uncertainty principle, already 
evident, will be discussed further in Sect. 12. 
\newline
Finally, consider a normalised distribution (\ref{1.1}) constructed
with the help of a complex valued function $\rho(f)$,
whose real and imaginary parts are contained in the interval
$0\le f\le 1$,
\begin{equation} \label{1.10}
\rho(f)=\rho_1(f)+i\rho_2(f),
\end{equation}
\begin{equation} \label{1.11}
\int_0^1\rho(f)df=A_1+iA_2.
\end{equation}
Any such distribution can be written as
as a sum of its real and imaginary parts
\begin{eqnarray} \label{1.12}
w(f)\equiv w_1(f)+iw_2(f) = \quad \quad \quad \quad \quad \quad \quad \quad \quad \quad \quad \quad \quad \quad \quad \quad \\
\nonumber
 \frac{A_1^2 (\rho_1(f)/A_1)+A_2^2 (\rho_2(f)/A_2)}
{A_1^2+A_2^2}+iA_1A_2\frac{\rho_2(f)/A_2-\rho_1(f)/A_1}{A_1^2+A_2^2},
\end{eqnarray}
 normalised to unity and zero, respectively,
\begin{equation} \label{1.13}
\int_0^1 w_1(f)df=1, \quad \int_0^1 w_2(f)df=0.
\end{equation}
It is readily seen that $w_1(f)$ is a is a proper distribution provided both 
$\rho_1$ and $\rho_2$ do not change sign for $0\le f \le 1$, 
while ,
$w_1(f)$ must alternate and can never be a valid probability distribution. 
For and improper $w_1(f)$
one can expect to obtain anomalously large values of
\begin{equation} \label{1.14}
Re\la f \ra = \int_0^1 f w_1(f)df
\end{equation}
when both normalisation integrals  $A_1$ and $A_2$ are small.
For example, for
\begin{equation} \label{1.15}
\rho(f)=\rho^{\epsilon_1}(f)+i\rho^{\epsilon_2}(f)
\end{equation}
we have
\begin{equation} \label{1.16}
Re\la f \ra = \frac{1}{2}-\frac{1}{2\pi} \frac {\epsilon_1 +\epsilon_2}
{\epsilon_1^2 +\epsilon_2^2}
\end{equation}
which can indeed take very large positive
and negative values in the vicinity of $\epsilon_1\approx \epsilon_2 \approx 0$,
whereas for $|\epsilon_1|,|\epsilon_2| > 1$, where $w_1(f)$
in non-negative, $Re\la f \ra$ remains positive and bounded by $1$.
\newline
In summary,
 the use of proper probability distributions largely relies
on interpreting the mean and variance as the location
and width of the region which contains physically 
significant values of the random variable $f$. 
 An improper distribution cannot, in general, be used for this purpose.
Analysis of this Section may seem an exercise of little practical importance,
as it is not immediately clear under which circumstances a
 physical quantity may be described by an 
 alternating let alone 
  complex-valued
 alternating distribution. We will, however, show that averages associated
 with such improper distributions naturally arise when one
 attempts  to obtain a answer to a question conventionally
 not answered by quantum mechanics, such as 
 determining the slit take by a particle in a diffraction
 experiment, or obtaining the value of a physical variable
 without perturbing the particle's motion.
 But first we consider the low accuracy limit of a 
 purely classical measurement.

 \section{Inaccurate classical measurements}
Consider a classical meter with a pointer position 
$f$ and a momentum $\lambda$ coupled to a one-dimensional
particle moving in a potential $V(x)$. The Hamiltonian 
for such a system is 
\begin{equation} \label{2.1}
H(p,x,\lambda)=p^2/2m+V(x)+ \beta (t)\lambda A(p,x)
\end{equation}
where the switching function $\beta(t)$ determines the 
strength of the coupling and $A(p,x)$ is the variable 
to be measured. It is readily seen that if the (conserved)
momentum  and the initial position of the meter are put to zero,
the pointer position time $t$ is given by
\begin{equation} \label{2.2}
f(t)=\int_0^t \beta(t')A(p,x)dt'
\end{equation}
so that the meter monitors the value of the functional in the r.h.s. of Eq.(\ref{2.2})
on the trajectory $\{p(t'),x(t')\}$ which is unaffected by the 
measurement. Let us assume in addition, that at $t=0$
the initial momentum and position of the particle, $P$ and $X$, are
not known precisely, but rather are random quantities distributed with 
the probability density $w(P,X)$. Since each trajectory is uniquely labeled by the values $(P,X)$, the value of the functional
(\ref{2.2}) is a random variable with a probability distribution 
\begin{equation} \label{2.3}        
w(f)= \int dP dX \delta(f-\int_0^t A(p,x)dt')w(P,X)\equiv <\delta(f-\int_0^t A(p,x)dt')>_{w}.
\end{equation}
and the first two moments of the form
\begin{equation} \label{2.3a}
<f>\equiv \int df f w(f) = \int_0^t dt' <A(t')>_{w}
\end{equation}
and
\begin{equation} \label{2.3b}        
<f^2>\equiv \int df f^2 w(f) = 2\int_0^t dt' \int_0^{t'} dt'' <A(t')A(t'')>_{w},
\end{equation}
where, as in Eq.(\ref{2.3}), $<...>_{w}$ denotes an average 
over all initial values $P$ and $X$.
\newline
Consider next a meter whose final position can be determined accurately,
but whose initial position is uncertain and is distributed around
$f=0$ with a normalised probability density $G(f)$ with a zero
expectation value and a known variance.
The two sources of uncertainty
result in  a simple convolution
formula for the normalised distribution $W(f)$ of the meter's readings,
\begin{equation} \label{2.3c}
W(f)=\int G(f-f')w(f')df'.
\end{equation}
Evaluating the generating function
\begin{equation} \label{2.5a}
 \la \exp(-i\lambda f)\ra _W =
\int \exp(-i\lambda f)W(f)df
\end{equation}
for the 
$n$-th moment of $W(f)$ we find
\begin{equation} \label{2.4}
\la f^n \ra_W = 2\pi i^n \partial
_{\lambda}^n\{\tilde {G}(\lambda)\tilde{w}(\lambda)\}
|_{\lambda=0}
\end{equation}
where the tilde denotes the Fourier transform, e.g.,
\begin{equation} \label{2.5}
\tilde {G}(\lambda)\equiv (2\pi)^{-1} \int \exp(-i\lambda f)G(f)df,
\end{equation}
$\partial{\lambda} ^n$ is the $n$-th derivative with respect to $\lambda$
and we used the  convolution property 
\begin{equation} \label{conv}
 \tilde{W}(\lambda)=
\tilde {G}(\lambda)\tilde{w}(\lambda).
\end{equation}
Applying the Leibniz product rule to the derivative (\ref{2.4}) 
yields
\begin{equation} \label{2.5c}
\la f^n \ra_W = \sum_{k=1}^n C^n_k \la f^n \ra_G\la f^n \ra_w,
\end{equation}
where $C^n_k$ are the binomial coefficients and, in particular, ($\la f\ra_G$=0)
\begin{equation} \label{2.6}
\la f \ra _W = \la f \ra_w,
\end{equation}
\begin{equation} \label{2.7}
\la f^2 \ra _W = \la f^2 \ra_G+\la f^2 \ra_w
\end{equation}
Next we ask what, if anything, can be learned about the mean and the variance 
of $w(f)$ in the low accuracy limit, when
the initial pointer position becomes highly uncertain ,
\begin{equation} \label{2.8}
G(f) \rightarrow \alpha^{-1} G(f/\alpha), \quad \alpha \rightarrow \infty.
\end{equation}
In this case $W(f)$ is a very broad distribution with the mean equal 
to that of $w(f)$ and a large variance
$D_W\equiv(\la f^2  \ra_W-(\la f \ra _W))^{1/2} \approx \alpha$.
The value $\la f \ra_w$ can be obtained by taking the average $\la f \ra_N$
of $N$
consecutive measurements. 
Since the variance of $\la f \ra_N$ is given by
$D/N$, so that to determine the value of $\la f \ra_w$
to an accuracy $\delta <<1$, one would require 
a very large number of measurements
\begin{equation} \label{2.4c}
N >> \alpha^2.
\end{equation}
Similarly, if $G(f)$ is known to a sufficient accuracy, one 
can use Eq.(\ref{2.7}) to determine the second moment of $w(f)$ and,
therefore, the original variance of $f$. Note, however,
that an accurate determination would require an even larger,
$N >> \alpha^4$, number of observations.
\newline
In summary, a classical meter with an increasingly
uncertain initial position is rendered impractical because, although
its readings contain the information about the mean
and variance of the measured variable,
the number of trials required for its 
extraction becomes prohibitively large.
\section {Quantum meters and measurements}
Consider next a similar measurement in the quantum case.
A detailed analysis of quantum meters has been given in \cite{SR1,SR2}
and here we will limit ourselves to only a brief 
discussion required for further development.
The  Schroedinger equation describing a system coupled to a von Neumann-like \cite{FOOT1} meter is
 ($\hbar =1$)
\begin{equation} \label{3.1}
i\partial_t |\Psi(t|f)\ra = [\hat{H} - i\partial_f \beta(t)\hat
{A}|\Psi(t|f)\ra.
\end{equation}
Initially, the system is prepared in some state $|\Psi_0\ra$, and the
meter position is set to zero,
\begin{equation} \label{3.2}
|\Psi(t=0|f)\ra=\delta(f)|\Psi_0\ra,
\end{equation}
Note that the Heisenberg's uncertainty principle prevents one from setting
the meter position to zero as well, which can be seen as the cause 
of the perturbation produced by the measurement \cite{SR1}.
After the measurement, at the time $t$, the state of the system is described
by the density operator
\begin{equation} \label{3.3}
\hat{\rho}=\int df |\Psi(t|f)\ra \la \Psi(t|f)|.
\end{equation}
If this state is purified, i.e., after the measurement the system
is post-selected in some state $|\Psi_1\ra$, the probability amplitude $\Phi(f)$
to obtain the meter reading $f$ is given by
\begin{equation} \label{3.4}
\Phi(f)= \la\Psi_1|\Psi(t|f)\ra.
\end{equation}
A useful representation for $\Phi(f)$ is obtained by solving 
Eq.(\ref{3.1}) by the Fourier transform,
\begin{eqnarray} \label{3.5}
\Phi(f)=
(2\pi)^{-1} \int d\lambda \exp(if\lambda)
\la \Psi_1| \hat{U}_{\lambda}|\Psi_0\ra/\la \Psi_1 |\hat{U}_{0}|\Psi_0\ra
\\
\hat{U}_{\lambda} \equiv \exp(-i\int_0^t [\hat{H}+\lambda \beta(t')
\hat{A}]dt'].
\end{eqnarray}
The measurement amplitude (\ref{3.5}) can be related to the value of the functional (\ref{2.2}) in the following way.
Although there is no unique trajectory, as in the
classical case, the transition amplitude between the initial
and the final states of the system in the absence of the meter can be written as a sum over the virtual
paths traced by the variable $A$,
\begin{eqnarray} \label{3.6}
\la \Psi_1|\exp(-i\hat
{H}t)|\Psi_0\ra = lim_{N\rightarrow \infty}\sum_{k_{1},k_{2},...k_{N}}
\la \Psi_1|\prod_{j=1}^N |a_{k_j}\ra 
\\
\nonumber
 \la a_{k_j}|\exp(-i\hat
{H}t/N)||a_{k_{j-1}}\ra \la a_{k_{j-1}}||\Psi_0\ra
\\
\nonumber
\equiv \sum_{[a]} \la \Psi_1|\Phi[a]\ra
\end{eqnarray}
where, as in the following, $a_k$ and $|a_k\ra$ are
the eigenvalues and eigenvectors of the variable of interest $\hat{A}$,
\begin{equation} \label{3.7}
\hat{A}|a_k\ra = a_k|a_k\ra.
\end{equation}
\newline
It can be shown \cite{SR1} that, in the presence of the meter, $\Phi(f)$ in Eq.(\ref{3.5})
is given by the restricted path sum 
\begin{equation} \label{3.8}
\Phi(f)= \sum_{[a]} \delta(f-\int_{0}^t \beta(t') a(t') dt')
\la\Psi_1|\Phi[a]\ra.
\end{equation}
The generalisation of Eq.(\ref{2.3}) to the quantum case is,
therefore, straightforward: a quantum pointer may be shifted
by an amount $f$ if among the paths contributing to the transition
some give value $f$ to the functional $F[a]= \int_{0}^t \beta(t') a(t') dt'$. 
The probability amplitude $\Phi(f)$ for the reading to occur is found by summing
the amplitude $\la \Psi_1|\Phi[a]\ra$ over all such paths.
The support of $\Phi(f)$ (i.e., the set of $f$ such that $\Phi(f)\ne 0$)
yields, therefore, the range of the values of $f$ which contribute 
to the transition.
It is also obvious that $\Phi(f)$ is a complex amplitude distribution
whose normalisation integral is the unperturbed transition amplitude,
\begin{equation} \label{3.8a}
\int \Phi(f) df = 
\la\Psi_1|\exp (-i\hat{H}t)|\Phi_0\ra.
\end{equation}
Since there are no {\it apriopi} restrictions on the sign of either $Re\Phi(f)$
or $Im\Phi(f)$, we cannot, as discussed in Sect.2(c), in general determine which values
within the support of $\Phi(f)$
contribute to the transition, in particular, when $\Phi(f)$ is of order of unity 
and $\la\Psi_1|\hat{U}(t)|\Phi_0\ra$ is very small.
This is, in essence, the Feynman's uncertainty principle.
One exception is the classical limit, in which a highly oscillatory $\Phi(f)$
has a stationary region near the classical value $f=f^{class}$, 
which is the only contributor to the integral (\ref{3.8a}) \cite{Spart}.
\newline
As in Section 3 we proceed with a discussion of a meter whose initial position 
so that the initial meter state in the position representation  is no longer
a $\delta$-function but rather some $G(f)$, with a finite width is $\Delta f$.
Then the amplitude 
$\Psi(f)$ to obtain the reading $f$ 
for the system post-selected in the state $|\Psi_1\ra$ can be written 
is a convolution  \cite{SR1}
\begin{equation} \label{3.9}
\Psi(f) = \int G(f-f')\la\Psi_1|\Phi(t|f')\ra df'.
\end{equation}
which is similar to  Eq.(\ref{2.6})
with the important difference that it relates probability amplitudes
rather than the probabilities themselves. If $\delta f$ is small,
we find the probability to obtain a reading $f$
$\rho(f) \approx |\Phi(f)|^2$
 so that an accurate meter measures the value of $F[a]$, and may
be used to evaluate the centroid and the width of the range of $f$
values which contribute to the transition amplitude between the states
$|\Psi_0\ra$ and $|\Psi_1\ra$. There is, however, a price. 
A measurement perturbs the system, whose
state
after a measurement yielding $f$ results not equal to that without 
a meter,
\begin{equation} \label{3.11}
\int G(f-f')\Phi(f')df' \ne \la \Psi_1 |\exp (-i\hat{H}t)|\Psi_0\ra
= \int\Phi(f')df',
\end{equation}
where the last equality is obtained by integrating Eq.(\ref{3.8}).
The perturbation can be minimised by choosing $G(f)$ so broad
that it can be replaced by a constant, making the l.h.s.
of Eq.(\ref{1.5}) proportional to $\exp (-i\hat{H}t)|\Psi_0\ra$
with an unimportant overall factor.
Whereas an uncertainty in the classical meter's initial position is clearly
undesirable, a similar uncertainty in the quantum case has the advantage
of reducing the perturbation a measurement produces on the measured system.
One then wishes to know what kind of information about
a quantum system can be obtained without affecting its evolution.
 
\section {Inaccurate quantum measurements, weak values
and negative probability}
Consider next the moments of a probability distribution for the meter's
readings,
\begin{equation} \label{4.1}
\rho(f)=|\Psi(f)|^{2}
\end{equation}
where $\Psi(f)$, given by Eq.(\ref{3.9}),
 is an complex valued and, possibly, alternating
amplitude distribution. Representing 
$\Psi(f)$ as a Fourier integral, calculating the generating function 
$\la \exp(-i\lambda f)\ra_{\rho}$ and expanding the result in the powers
of $k$ yields
\begin{equation} \label{4.2}
\la f^n \ra \equiv \int f^n \rho(f)df/\int \rho(f)df=
\int  i^n\frac {\partial^n_{\lambda}
\tilde{\Psi}(\lambda)}{\tilde{\Psi}(\lambda)}
|\tilde{\Psi}(\lambda)|^2 
d\lambda
/\int |\tilde{\Psi}(\lambda)|^2 d\lambda
\end{equation}
Recalling that $i^n\partial^n_{\lambda}\tilde{\Psi}(\lambda)/
\tilde{\Psi}(\lambda)=\int f^n \exp(-i\lambda f)\Psi(f) df/
\int\exp(-i\lambda f)\Psi(f) df$ shows that
the moments of the proper probability distribution $\rho$
can be expressed via the moments
of a family of improper complex distributions
$\{\exp(-i\lambda f)\Psi(f)\}$ for all $-\infty < \lambda < \infty$.
\newline Calculating the Fourier transform of the convolution (\ref{3.1}) %
for the first two moments we 
obtain 
\begin{equation} \label{4.4}
\la f \ra = i\int \tilde{G}^*
\tilde{\Phi}^*[\tilde{G}'\tilde{\Phi}
+\tilde{G}\tilde{\Phi}']d\lambda / \int |\tilde{G}|^2
|\tilde{\Phi}|^2 d\lambda
\end{equation}
and
\begin{equation} \label{4.5}
\la f^2 \ra  = -\int \tilde{G}^*
\tilde{\Phi}^*[\tilde{G}''\tilde{\Phi}+2\tilde{G}'\tilde{\Phi}'
+\tilde{G}\tilde{\Phi}'']d\lambda / \int |\tilde{G}|^2
|\tilde{\Phi}|^2 d\lambda.
\end{equation}
 It is natural to choose the initial state of the meter to be 
a real even function with a width of order of unity
, $G(f)^*=G(f)$, $G(f)=G(-f)$, e.g., 
a Gaussian, so that
\begin{equation} \label{4.6}
\tilde{G}(\lambda)=\tilde{G}^*(\lambda), \quad
\tilde{G}(\lambda)=\tilde{G}(-\lambda).
\end{equation}
As in the classical case, the accuracy of the 
quantum measurement can be reduced by increasing the
initial uncertainty of the pointer's position,
\begin{equation} \label{4.7}
G(f)\rightarrow G(f/\alpha), \quad
\tilde{G}(\lambda) \rightarrow \tilde{G}(\alpha \lambda),
\quad \alpha \rightarrow \infty.
\end{equation}
As the uncertainty increases, the Fourier transform
$\tilde{G}(\alpha \lambda)$ becomes sharply peaked 
around $\lambda = 0$, and the integrals
(\ref{4.4}) and (\ref{4.5}) can be evaluated by expanding
the terms containing $\tilde{\Phi}$ in the Taylor series.
Estimating
\begin{equation} \label{4.8}
\int G^*(\alpha \lambda) \partial^n_{\lambda}
G(\alpha \lambda)\lambda^m d\lambda
= \alpha^{n-m-1}\int G^*(z) \partial^n_{z}
G(z)z^m dz = O(\alpha^{n-m-1})
\end{equation}
and retaining the leading terms in $\alpha ^{-1}$, we obtain
\begin{equation} \label{4.9}
\la f \ra  = Re \bar{f} + O(\alpha^{-1}),
\end{equation}
and 
\begin{equation} \label{4.10}
\la f^2 \ra = \alpha^2 \frac{\int z^2G(z)^2 dz}{\int G(z)^2 dz}
+C(Re \bar{f^2}-|\bar{f}|^2)+|\bar{f}|^2+O(\alpha^{-1}).
\end{equation}
where the factor $C$ is given by
\begin{equation} \label{4.12}
C\equiv \frac{\int z^2 \tilde{G} \tilde{G}''dz}
{\int \tilde{G}^2dz} -\frac{\int z^2 \tilde{G}^2 dz}
{\int \tilde{G}^2dz}
 \frac{\int \tilde{G} \tilde{G}''dz}
{\int \tilde{G}^2dz}.
\end{equation}
and we have introduced the notation $\bar{f^n}$ for the $n$-th moment 
of the complex valued
amplitude distribution $ \Phi(f)$ defined in Eq.(\ref{3.8}),
\begin{equation} \label{4.11}
\bar{f^n}\equiv 
 \int f^n \Phi(f) df/\int \Phi(f) df =
 (i)^n \partial^n_{\lambda}
 \tilde{\Phi}(0)/\tilde{\Phi}(0),
\end{equation}
Expressions similar to Eq.(\ref{4.9}) have earlier been 
obtained in  \cite{Ah1,AhBOOK} for a weak von Neumann measurement
and in \cite{SB} for the quantum traversal time.
\newline
In summary, Eqs. (\ref{4.9}) and (\ref{4.10}), obtained here for an 
inaccurate von Neumann-like measurement of Sect.4, constitute
a  more general illustration of the uncertainty principle.
Whenever probability amplitude for a variable $f$
is obtained by smearing the amplitude for a variable
$f'$ with a broad envelope function so that the coherence
between different values of $f'$ is not destroyed, evaluating  $\la f \ra$ and $\la f^2 \ra$ 
does not, in general, reveal the mean and variance
obtained in an accurate measurement of $f'$.
Rather, the values $\la f \ra_w$ and $\la f'^2 \ra_w$ in the classical
 Eqs.(\ref{2.6}) and (\ref{2.7}), are replaced by 
 the weak value $Re\bar{f}$  and 
 a
complicated combination of 
$\bar{f}$ and $\bar{f^2}$, respectively.
Since there is no restriction on the 
phase of $\Phi(f)$ in Eq.(\ref{4.11}), interpretation of these these quantities
as averages
requires the concept of negative probability.
As a result the information about the values 
  of $f'$ which contribute to the transition,
   may be 'scrambled' by averaging 
with an improper alternating distribution. 

\section {Where was the particle half way through a transition?}
This unhelpful property of the weak values $\bar{f}$ is most easily illustrated
on the double-slit diffraction experiment. Consider 
a point on the screen such that
the amplitudes to reach it via the slit $1$ and the slit $2$ 
are $\Phi(1)=1$ and $\Phi(2)=-1+\epsilon$, respectively,
and attempt to determine the mean slit number using 
Eq.(\ref{4.9}). The variable $f$ can only takes two values 
$1$ and $2$, and the integrals in Eq.(\ref{4.2}) are be replaced
by sums, which gives
\begin{equation} \label{5.1}
\bar{f}=2-1/\epsilon
\end{equation}
For $\epsilon=0.1$ Eq.(\ref{5.1}) yields $\bar{f}=-8$
and it is difficult to interpret the notion that an 
electron passes on average through the slit number
$-8$ as anything
other than a failure of our measurement procedure.
\newline
Analysis of Section 5  allows to apply exactly the same
reasoning to a more conventional von Neumann measurement
of the type considered in \cite {Ah1}. Consider a two 
level system with a zero Hamiltonian $\hat{H}=0$
and the 'position'  operator  (c.f. 
the position operator $\hat{x}=\int |x\ra x \la x|dx$ for a particle
in one spatial dimension)
\begin{equation}
\hat{A}=|1\ra \la 1|+|2\ra 2 \la 2|,
\end{equation}
 prepared and post-selected in the states
($N_0$ and $N_1$ are the normalisation constants)
\begin{equation} \label{5.2}
|\Psi_0\ra = N_0(|1\ra+|2\ra)\quad and \quad |\Psi_1\ra 
= N_1(|1\ra-(1-\epsilon)|2\ra),
\end{equation}
respectively.
To determine which state the system was at, say,
$t/2$ we may employ a von Neumann meter
with a Gaussian initial state
\begin{eqnarray} \label{5.3c}
G(f)=\exp(- f^2/\alpha^2).
\end{eqnarray}
Equation (\ref{3.6}) shows that for $\hat{H}=0$ only two paths
connecting the initial and final states, $a(t)=1$ and 
$a(t)=2$, have non-zero  probability amplitudes
\begin{eqnarray} \label{5.3}
\Phi(1)=\la \Psi_1|1\ra \la 1| \Psi_0\ra=1
\\ \nonumber
\Phi(2)=\la \Psi_1|2\ra \la 2| \Psi_0\ra= \epsilon - 1,
\end{eqnarray}
respectively (see Fig.1).
To obtain the system's position at $t/2$ we choose
the switching function in Eq.(\ref{3.3}) to be 
$\beta(t')=\delta(t'-t/2)$ which yields 
\begin{eqnarray} \label{5.4c}
\Phi(f)=\delta(f-1)-(1-\epsilon)\delta(f-2)
\end{eqnarray}
and the average pointer position is given by
\begin{eqnarray} \label{5.4}
\la f \ra = \frac {1+2(1-\epsilon)^2-3(1-\epsilon)\exp(-1/2\alpha^2)}
{1+(1-\epsilon)^2-2(1-\epsilon)\exp(-1/2\alpha^2)}
\end{eqnarray}
for an arbitrary resolution $\Delta f = \alpha$.
In the high accuracy limit, $\alpha \rightarrow 0$, 
 for $\epsilon<<1$ we obtain
$\la f \ra \approx 1.5$ which indicates that the observed
system would be found in each of the two states
with equal probability.
The 
probability of transition to the state $|\Psi_1\ra$ would, however, 
be altered by the measurement,
\begin{eqnarray} \label{5.4b}
P=N_0^2 N_1^2(1+(1-\epsilon)^2)  \ne N_0^2 N_1^2\epsilon^2 = |\la \Psi_1|\Psi_0 \ra |^2.
\end{eqnarray}
To keep the transition probability unchanged
we may apply a highly inaccurate meter with $\alpha \rightarrow \infty$. 
For $\epsilon << 1$ the initial and final states are
nearly orthogonal and, based on the discussion at the
end of Sect.2, we expect the weak value obtained as
$\alpha \rightarrow \infty$ to be without a direct relation
to the two actual positions $f=1$ and $f=2$, which contribute
to the transition.
Indeed, in this limit we recover
Eq.(\ref{5.1}) and for $\epsilon=0.1$ again find 
the measured mean position $\la f \ra=-8$. 
The dependence of $\la f \ra$ on the resolution
$\alpha$ and the parameter $\epsilon$ is shown in Fig.2.
Finally we note that, in a similar way,
a transition amplitude for a system
with three or more discrete states can be mapped onto
a diffraction experiment with three or more slits.
We will return to this analogy in Sect.7.

\section{Weak measurements without post-selection.}

Until now we have assumed that the system is post selected
after a measurements in a known state $|\Psi_1|\ra$
so that the meter reading are sampled only if it is found in $|\Psi_1|ra$,
and discarded otherwise.
If the system's final is not controlled and all the reading are
kept, the results (\ref{4.9}) and
(\ref{4.10}) must be averaged further  with the probabilities 
$P_m$ to find the system in the state $|m\ra$ belonging to some
orthonormal set. As our measurement is weak, $P_m$ are essentially
the same as in the absence of the meter,
\begin{eqnarray} \label{5.5}
P_m=|\la m|\exp(-i\hat{H}_0t)|\Psi_0\ra|^2
\end{eqnarray}
so that we have
\begin{eqnarray} \label{5.6}
\la \la f^n \ra \ra \equiv \sum_m P_m \la f^n \ra_m.
\end{eqnarray}
Here and in the following extra angular bracket 
denotes average with $P_m$ and 
the (previously suppressed) subscript $m$ on and average or distribution
indicates that it has been  evaluated for the final state
for $|\Psi_1\ra = |m\ra$.
In particular, for the mean we have
\begin{eqnarray} \label{5.7}
\la \la f \ra \ra = Re \sum_m P_m \bar{f}_m \equiv  
Re \la \bar{f}\ra.
\end{eqnarray}
where newly introduced average  $\la \bar{f}\ra$
is the mean calculated with the distribution
\begin{eqnarray} \label{5.8}
\la \Phi (f) \ra  \equiv \sum_m P_m \Phi_m(f) =
(2\pi)^{-1} \int d\lambda \exp(if\lambda)
\la \Psi_0| \hat{U_0}^{-1} \hat{U}_{\lambda}|\Psi_0\ra
\equiv w_1(f)+iw_2(f)
\end{eqnarray}
which is a weighted sum of improper distributions $\Phi_m(f)$
and is, for this reason, itself an improper distribution.
Recalling the relation (\ref{2.4}) between the moments
and the Fourier transform of a distribution, Eq.(\ref{3.5}), and
using the perturbation theory to expand the evolution
operator $\hat{U}_{\lambda}$ in powers of $\lambda$, we 
find (c.f. the classical Eqs. (\ref{2.3a}) and (\ref{2.3b}))
\begin{eqnarray} \label{5.9}
\la \bar{f} \ra = \int_0^t \beta(t') 
\la \Psi(t')|\hat{A}|\Psi(t')\ra dt' \quad \quad \quad \quad \quad \quad \quad \quad \quad \quad \quad \quad \quad \quad 
\\
\la \bar{f^2} \ra = \int_0^t dt'' \int_0^{t''} \beta(t')\beta(t'')
\la \Psi(t'')|\hat{A}
\exp[-i\hat{H}_0(t''-t')]\hat{A}|\Psi(t')\ra, 
\end{eqnarray}
where $|\Psi(t')\ra \equiv \exp(-i\hat{H}_0t')|\Psi_0\ra$.
\newline
To calculate $\la \la f^2 \ra \ra$ we will require
simple sum rules, resulting from the Hermitian nature
of the operator $\hat{A}$,
\begin{equation} \label{5.10}
I(\lambda) \equiv \sum_m \la \Psi_0|\hat{U}^{-1}_{\lambda}|m\ra
\la m|\hat{U}^{-1}_{\lambda}|\Psi_0\ra =1.
\end{equation}
Calculating $\partial_{\lambda}I(0)$ and $\partial^2_{\lambda}I(0)$
and using Eq.(\ref{3.8}) we find
\begin{eqnarray} \label{5.11}
Im\la \bar{f} \ra =0
\\
Re \la \bar{f^2} \ra = \sum_m P_m |\bar{f}_m|^2
\end{eqnarray}
The first of these relations confirms that $\la \bar{f} \ra$
is real, as is already evident from Eq.(\ref{5.9}), and the 
second helps us average Eq.(\ref{4.10}) to obtain
\begin{equation} \label{5.12}
\la \la f^2 \ra \ra = \alpha^2 \frac{\int z^2G(z)^2 dz}{\int G(z)^2 dz}
+Re \la \bar{f^2} \ra
\end{equation}
Equations (\ref{5.7}) and (\ref{5.12}) are the central result of this
Section. They have the same form as the classical equations
(\ref{2.6}) and (\ref{2.7}) insofar as the l.h.s. of Eq.(\ref{5.7}) and
the second term in Eq.(\ref{5.12}) are the first two moments
of the same distribution $w_1$ in Eq.(\ref{5.8}).
However,
owing to the inaccuracy of the measurement, 
 $w_1$ can, in general, change sign
 and one must exercise caution when using these averages.
 For example,  Sect.2, it has been shown that for such distributions it is possible
 to have 
 \begin{equation} \label{5.14}
 Re \la \bar{f^2} \ra = (Re \la \bar{f} \ra)^2,
 \end{equation}
 while $f$ remains a distributed, rather than a sharply
 defined quantity. 
Indeed, the relation (\ref{5.14}) will always take place
for $\hat{A}$ and $\hat{H}_0$ such that, regardless of the 
value of $\lambda$, $\hat{U}_{\lambda}$ evolves the
initial state $\Psi_0\ra$ into the same final state $\Psi_1\ra$,
so that in Eq.(\ref{5.8}) 
\begin{equation} \label{5.15}
\la \Psi_0| \hat{U_0}^{-1} \hat{U}_{\lambda}|\Psi_0\ra
= \exp[i\phi (\lambda)]\equiv S(\lambda)
\end{equation}
 where $\phi (\lambda)$ is a real phase, as required by
 the unitarity. As a result we have 
  \begin{equation} \label{5.16}
  Re(\la \bar{f^2} \ra)=Re[-S^{-1}(0)S''(0)]=\phi '(0)^2=
  (Re[iS^{-1}(0)S'(0)])^2= (\la \bar{f} \ra)^2.
  \end{equation}
while, apparently, $Re[(2\pi)^{-1}\int d \lambda \exp(i\lambda f) \exp
[i\phi (\lambda)-i\phi(0)] \ne \delta (f-\la \bar{f^2} \ra)$.
\newline
In summary, without post-selection 
one recovers the classical Eqs.(\ref{2.6}) and (\ref{2.7}),
with the important difference that both the mean and the variance
are obtained with an (possibly) improper distribution $w_1(f)$.
Also, as shown in Sect.2, anomalously large weak values are likely
to occur for nearly forbidden transitions, whose probability
is quite small. For this reason they do not contribute
if the final state of the system is not controlled and
an average is taken over all possible final states.
Next we give further examples of (\ref{5.7}) and (\ref{5.12}),
starting with the conventional von Neumann measurement.

\section{Weak von Neumann measurements as a special case}

An important special case of von Neumann-like measurements
described in Sect. 5 are  impulsive von Neumann measurements,
already briefly discussed in Sect.4,
 whose
weak limit has been first analysed by Aharonov {\it et al}
in Refs.\cite {Ah1}.
The purpose of such a measurement is to establish the value $f$
of a variable $\hat{A}$ with a discrete spectrum
$\{a_k\}$ at some intermediate time $t_0$
for a system initially prepared in a state $|\Psi_0\ra$ and
then post-selected in a final state $|m\ra$
The the probability amplitude $\Phi(f)$ is, in this case,
the net amplitude on all virtual egenpaths in Eq.(\ref{3.6}), which at 
$0\le t_0 \le t$  pass through the value $f=a_k$.
Thus putting in Eq.(3.1) 
\begin{eqnarray} \label{8.1}
\beta(t')= \delta(t'-t_0)
\end{eqnarray}
and evaluating the Fourier transform (\ref{3.5})
we obtain
\begin{eqnarray} \label{8.2} 
\Phi(f)=\la m|\Psi(t)\ra^{-1}\sum_k \delta(f-a_k)
\la n|a_k\ra  \la a_k|\Psi(t_0)\ra
\end{eqnarray}
Where $|n\ra$ is the state obtained by evolving 
$|m\ra$ back to the time $t_0$, $|n\ra \equiv
\hat{U}^{-1}(t-t_0)|m\ra$, and $|\Psi(t)\ra\equiv 
\hat{U}(t_0)|\Psi_0\ra$.
Thus $\Psi (f)$ is, as expected, a complex valued
distribution, whose support coincides with  
spectrum of the operator $\hat{A}$. 
In general, the distribution is an improper one,
as the real and imaginary parts of the complex coefficients
$\la n|a_k\ra  $ and $  \la a_k|\Psi(t_0)\ra$ 
which multiply the $\delta$-functions can take either sign.
\newline
If no post-selection is made, the distribution
(\ref{8.2}) needs to be averaged over all final states
$|m\ra$ and Eq.(\ref{5.7}) gives
\begin{eqnarray} \label{8.2a}
\la \Phi(f)\ra=
\sum_k \delta(f-a_k)
 |\la a_k|\Psi(t_0)\ra|^2
\end{eqnarray}
Now the weight multiplying the $\delta$-functiions
are strictly non-negative  and, unlike $\Phi(f)$ the averaged distribution
is a proper one.
What is more, $\la \Phi\ra(f)$ coincides with
the probability distribution obtained for and accurate
'strong' measurement of the variable $\hat{A}$
in a state $|\Psi(t_0)\ra$.
\newline
In summary, for weak von Neumann measurements 
without post selection we recover the classical Eqs. (\ref{5.7}) and (\ref{5.12})
which allow to extract  the mean and variance,
obtained in accurate measurements, from a large sample
of weak results.
Also, finding $\la\bar{f^2} \ra = \la\bar{f} \ra^2$
would, in this case, guarantee that the variable is 
sharply defined, i.e., that $|\Psi(t_0)|\ra$ is one of the 
eigenstates  of $\hat{A}$.
However, extending this  argument to the  
case when the measured quantity is not an instantaneous
value of an operator can lead to errors, as will be
shown in the next Section.

\section{Is the elastic collision time sharply defined?}
It is possible then that someone not familiar with
the analysis of Sect. 7 and implicitly assuming the values
(\ref{5.14}) obtained with a weak von Neumann-like meter to be proper
probabilistic averages, might incorrectly conclude that the value $f$
of a functional $F[a]$ is sharply defined, i.e., has 
a unique precise value.
One such example is the distribution of the elastic 
collision time studied by Baz' with the help
of a weakly coupled semiclassical Larmor clock \cite{Baz1,Baz2}.
In Baz' approach, a small constant magnetic field 
along the $z$-axis is created
in a sphere containing the target and a particle, described
in the distant past by an incoming plane wave $\exp(-ikr)$
is equipped with large nearly classical spin $j >> 1$
initially polarised along the $x$ axis. 
The spin rotates
for as long as the particle remains inside the sphere,
and after the collision the spin
of the outgoing particle is rotated
in the $xy$ plane. The mean collision time $\bar{\tau}$,
and its mean square $\bar{\tau^2}$ are then {\it defined} as
\begin{eqnarray} \label{5.17}
\bar{T}= (\omega j)^{-1} \bar{j_y}
\\ \nonumber
\bar{T^2}=(\omega j)^{-2} [\bar{j_y^2}-j/2]
\end{eqnarray}
where $\bar{j_y}$ and $\bar{j_y^2}$ are the expectation 
values of the spin's $y$-component and its square, 
respectively and $\omega$ is the Larmor frequency
A simple calculation shows that
\begin{equation} \label{5.18}
\bar{T^2}=(\bar{T})^2
\end{equation}
which led Baz' to conclude that 'for given energy $E$
and angular momentum $l$ the time interval during which
the colliding particles are inside a sphere of radius
$R$ is a sharply defined quantity' \cite{Baz2, BazBOOK}.
The matter was further discussed in Refs. \cite{LEV} and
and briefly mentioned in \cite {SC3}. 
The purpose of this Section is to show that 
for $j>>1$
$\bar{\tau}$ and $\bar{\tau^2}$ in
Eqs.(\ref{5.17})  are just the weak values
calculated for the traversal time functional \cite{SB}
($\theta_R(\vec{r})= 1 \quad for \quad r<R \quad and \quad 0 \quad otherwise$)
\begin{equation} \label{5.18a}
\tau [\vec{r}(.)]=\int_{-\infty}^{\infty} \theta_R (\vec{r})dt'
\end{equation}
which computes the net duration spent by a Feynman path $\vec{r}(t)$ inside
the sphere of the radius $R$ \cite{FOOT2} and that these values obey Eq.(\ref{5.16}).
Indeed, it can be shown \cite{SBOOK} that the final state of the clock's
spin, $|M_F\ra$, is just a superposition of rotations of 
its initial state $|M_I\ra$ around the $z$-axis by the angles
$\omega \tau$ each weighted
by the amplitude distribution $\Phi(\tau)$ with which the duration $\tau$
contributes to the collision. Thus, expanding in the eigenstates
$|m\ra$, $m=-j,...,j$ of the $z$-component of the spin, $\hat{j_z}$
we have
\begin{equation} \label{5.19}
\la m|M_F\ra =\int d\tau \Phi(\tau) \exp(-im\omega \tau) \la m|M_I\ra.
\end{equation}
which shows that the Larmor clock is similar to von Neumann like
meter of a kind described in Sect.4.
For a large spin polarised along the $x$-axis Baz' wrote
\begin{equation} \label{5.19a}
\la m|M_I\ra = C \exp(-m^2/2j),
\end{equation}
which restricts $|m| \le j^{1/2}$. 
The matrix $\la m'|\hat{j_y}| m\ra$
has two non-zero off-diagonal elements \cite{LAND}, 
$\la m+1|\hat{j_y}| m\ra
= -\la m|\hat{j_y}| m+1\ra =-i(j+m)^{1/2}(j-m+1)^{1/2}/2$.
 With the restriction on $|m|$, for a large $j$ 
 we 
 may write $\hat{j_y} \approx -ij \partial_m$
 so that in the continuous limit,
\begin{equation} \label{5.19b}
j\rightarrow \infty, \quad
\omega \rightarrow 0,\quad 
\omega j \rightarrow \infty \quad
\omega^2 j \rightarrow 0, 
\end{equation}
after introducing $\lambda \equiv m\omega$ we have
\begin{eqnarray} \label{5.20}
\bar{j_y^n}/j^n \omega^{n+1}\equiv \sum_{m,m'}
 \la M_F|m'\ra\la m'|\hat{j_y^n}| m\ra\la m|M_F\ra/j^n \omega^{n+1}
 = \quad \quad \quad \quad \quad \quad \quad \quad \quad \quad \quad \quad
 \\
 \nonumber
 \int d \lambda \exp(-\lambda^2/2\omega^2j)
 \tilde{\Phi}^*(\lambda) \partial^n_{\lambda} \{
 \exp(-\lambda^2/2\omega^2j)
  \tilde{\Phi}(\lambda)\}/ \int d \lambda \exp(-\lambda^2/\omega^2j)| \tilde{\Phi}(\lambda)|^2
\end{eqnarray}
in which we recognise Eqs.(\ref{4.2})  and  (\ref{4.7}) with $\alpha^2 = 1/\omega^2j
\rightarrow \infty $ and $\tilde{G}(\alpha \lambda)\equiv \exp(-\lambda^2/2\omega^2j)$
and Eqs. (\ref{5.17}) are seen to be equivalent to Eqs.(\ref{5.7}) and (\ref{5.12}).
Thus, for
 a small Larmor frequency the first of Eqs.(\ref{5.17})
gives the improper weak value of the traversal time Eq.(\ref{5.18a}).
\newline
Finally, for the functional  (\ref{5.18}) the evolution
operator $\hat{U}_\lambda = \exp[-i(\hat{p}^2/2m+V(r)+\lambda \theta_R(r))]$
contains an additional constant potential inside the sphere
of interest  which modifies the scattering phase $\phi(k)$,
so that (cf. Eq.(\ref{5.15})
\begin{equation} \label{5.22}
\hat{U}_\lambda \exp(-ikr) = \exp[i\phi(k,\lambda)]\exp(ikr)
\end{equation}
and therefore, according to Eq.(\ref{5.16}), 
\begin{equation} \label{5.23}
\bar{T^2}=Re\bar{\tau^2}=(\bar{\tau})^2= \bar{T}^2.
\end{equation}
For a rectangular potential,
$V(r)=\Omega \theta_R(\vec{r})$,
the traversal time amplitude distribution $\Phi(\tau)$ in Eq.(\ref{5.19})
and the weak value $\bar {\tau}$ vs. $\Omega$ are shown 
in Figs. 3a and 3b, respectively.
\newline
In summary, 
 the suggestion that the collision time has
a precise value in elastic scattering is shown to be incorrect.
Rather, Baz' result demonstrates that, for such a single-channel
collision, the real part of the traversal time amplitude,
$Re\{\Phi(\tau)/ \int \Phi(\tau) d \tau\}$ is a broad improper distribution with vanishing
variance. As was also observed by Baz' \cite {Baz3}, 
this is no longer true if a particle is post-selected
in one of several channels, e.g.,
for transmission across a potential barrier where both reflection and transmission
are possible. 

\section{Time delay in transmission and the phase time}

A different type of the time delay variable, not directly related
to the traversal time functional (\ref{5.18a}) or indeed to any other
functional of the particle's Feynman paths can be 
constructed as follows.
Consider a classical particle with a unit mass in one dimension crossing from left to right a
potential $V(x)$ which vanishes everywhere outside the 
region $-a<x<a$. Inside the region the particle will experience a time
delay or a speed up depending on whether $V(x)$ is a barrier or a well.
This time delay $\tau$ can be evaluated by taking a snapshot of the particle's
at some large time $t$ and comparing it with the position of a particle 
that has been moving freely along the trajectory 
with the same initial conditions. If the distance between the two 
is $x'$, we have ($p$ is the particle's initial momentum)
\begin{equation} \label{7.1}
\tau(p) = -x'/p
\end{equation}
which is positive (delay) if the particle lags behind, 
or negative (speed up) if it lies ahead of the free one. 
This is a measurement which differs from the one discussed
in Sect.4 in that the role of the pointer is
 played by the particle's own position, but a measurement 
nevertheless.
It is not surprising, therefore that a quantum extension
of such a procedure is a quantum measurement.
Initilally one represents a particle by a wavepacket
\begin{equation} \label{7.2c}
\Psi(x,t=0) = G(x)\exp(ipx) =\int A(k)exp(ikx)dk.
\end{equation}
 The transmitted part,
 is then given by
\begin{equation} \label{7.2}
\Psi^T(x,t) = \int T(k)A(k)exp(ikx - ik^2t/2)dk
\end{equation} 
where $T(k)$ is the transmission amplitude. 
Rewriting the Fourier transform (\ref{7.2}) as a convolution 
and neglecting the spreading of the wavepacket yields \cite{SMS}
\begin{equation} \label{7.5}
\Psi^T(x,t) = \exp(ipx-ip^2t/2)\int G(x-pt-x') \Phi_p (x') dx',
\end{equation}
where
\begin{equation} \label{7.6}
\Phi_p (x)\equiv (2\pi)^{-1}\exp(-ipx)\int T(k)\exp(ikx)dk.
\end{equation}
and
\begin{equation} \label{7.7}
\int \Phi_p (x)dx = T(p).
\end{equation} 
On sees that the transmitted
 wavepacket is constructed from the freely propagating envelopes
each shifted by $x'$ and weighted by the probability amplitude
$\Phi_p (x')$.
Associating with each spatial shift $x'$ a time delay $\tau$
with the help of Eq.(\ref{7.1}) shows 
that transmission of a particle with a momentum
$p$ involves not one but many time delays, whose amplitude
distribution is given by the Fourier  transform (\ref{7.6})
of $T(p)$. Moreover, observing transmitted particle at a location
$x$ amounts to measuring $\tau$ to the accuracy determined
by the coordinate spread of the particle's initial wavepacket.
\newline
To quantify the mean time delay associated with the 
latter one often chooses \cite{Rev1,Rev2,Rev3}
the shift of the centre of mass of the transmitted
pulse relative to that of the free propagation divided
by its mean velocity.
\begin{equation} \label{7.9}
<\tau (p)> \equiv p^{-1}<x-pt> = \int (x-pt) |\Psi^T(x,t)|^2dx/\int |\Psi^T(x,t)|^2dx, 
\end{equation}
i.e., the expectation value of the time delay for a particle with 
the momentum $p$ measured with the 'apparatus function' $G$
determined by the envelope of the pulse (cf. Eq.(\ref{3.9})).
Just as in the case of a von Neumann like
measurement (\ref{3.9}), an improvement in the accuracy increases
the 'perturbation' on the measured system, as a wavepacket narrow
in the coordinate space has a large momentum spread.
As a result, the transmission probability $P^T$ is not equal to that for
a plane wave with the momentum $p$, 
\begin{equation} \label{7.8}
P^T = \int |T(k)|^2|A(k)|^2 dk \ne |T(p)|^2.
\end{equation}
In order to minimise this perturbation one can choose
$A(k)$ so narrow that the inequality (\ref{7.8}) becomes an
approximate equality, and the envelope $G(x)$ becomes very broad.
As was shown in Sect.4, such a measurement 
is weak and the mean time delay is given by the real part of the improper weak
value
\begin{equation} \label{7.10}
<\tau (p)> \approx p^{-1}Re \bar{\tau}(p) \equiv p^{-1} Re \int x' \Phi_p(x')dx'
/ \int\Phi_p(x')dx'. 
\end{equation}
With the help of Eq.(\ref{7.7}) it is easy to show that Eq.(\ref{7.10}) can also 
be written as (cf. Eq.(\ref{4.9}))
\begin{equation} \label{7.11}
<\tau (p)> \approx p^{-1/2} Re[-i \partial_p ln T(p)]
= \partial_E \phi(p) \equiv \tau_{phase}, 
\end{equation}
where $\phi(p)$ is the phase of the transmission coefficient
$T(p)$, $T(p)=|T(p)|\exp[i\phi(p)]$.
Equation (\ref{7.11}) is the standard definition of the 'phase
time' \cite{Rev1, Rev2,Rev3}. 
The purpose of the above analysis has been to clarify
its origin as an weak value and relate it to some of its 
'anomalous' properties.
One such property property is that for tunnelling across a potential
barrier $\tau_{phase}$ predicts a speed up as if the classically 
forbidden region had been crossed infinitely fast. Indeed,
for tunnelling across
 a high rectangular barrier of a height $V > p^2/2$ and a width $a$ the transmission
coefficient can be approximated as (we neglect the pre-exponential factor)
\begin{equation} \label{7.12}
T(p) \approx \exp[-(V-p^2)a-ipa]
\end{equation}
so that
\begin{equation} \label{7.13}
\tau_{phase} \approx -a/p. 
\end{equation}
At first glance this result appears to contradict the relativistic
restriction that the speed of a particle or a photon may not exceed
the speed of light, but only if $\tau_{phase}$ is taken to be the time delay
in the classical sense.
In reality, it is just a spectacular example of an improper 
average lying outside the region of support of a continuous oscillating distribution
(\ref{7.6}) .
Indeed, as the barrier potential does not have bound states
and, therefore, poles in the upper half of the complex $k$-plane,
$\Phi_p(x')$ in Eq.(\ref{7.6}) vanishes for $x'>0$ so that only
positive time delays contribute to tunnelling of a particle
with a momentum $p$. Thus the causality is not violated and 
the 'anomalous' negative value (\ref{7.13}) simply indicates the possibility
that below the barrier destructive interference between the delayed
envelopes may produced a significantly reduced advanced pulse 
which builds up from their front
 tails (more details of this analysis can be found in \cite{SMS}). 
 For a zero-width barrier, $V(x)=\Omega \delta(x)$,
 the amplitude distribution $\Phi_p(x)$ in Eq.(\ref{7.6}) and the phase
 time (\ref{7.11}) vs. $\Omega$ in Figs. 4a and 4b, respectively.
 Note here the principal difference between $\tau_{phase}$ and the collision (traversal)
 time of the previous Section. While 
 the traversal time represented by the functional 
 (\ref{5.18a}) vanishes with the size of the region of interest ,
 $\tau_{phase}$ which relates to the poles of $T(k)$ in the complex 
 $k$-plane remaines finite for an infinitely narrow barrier.
 Thus the traversal time  and the phase time are essentially different 
 quantities which share the same classical limit  and Baz' assertion that the former
 is correct the latter is wrong \cite{BazBOOK}  cannot be sustained.

\section{The local angular momentum (LAM) }
A different example of a weak value is the local angular momentum,
designed and applied in \cite{LAM} to analyse elastic, inelastic 
and reactive differential cross-sections (DCS). 
Typically, several angular momenta contribute to the scattering
amplitude $f(\theta$, which is given by a coherent sum
over partial waves,
\begin{equation}\label{9.1}
f(\theta) \equiv (ik)^{-1/2}\sum_{J=0}^{\infty}
(J+1/2)P_J(\cos(\theta))S^J(E)
\end{equation}
where $\theta$ is the scattering angle,
$k$ is the wavevector,
 $J$ is the total angular momentum,
 $P_J(\cos(\theta))$ is the Legendre polynomial,
 and $S^J$ is the $S$-matrix element.
In order to estimate the angular momentum 
which contributes to a particular angle $\theta$ 
the authors of  \cite{LAM} suggested the quantity
with the units of angular momentum
\begin{equation}\label{9.2}
LAM (\theta) \equiv d \phi(\theta)/d\theta = Re[-i\partial_{\theta}
ln f(\theta)],
\end{equation}
which can be seen to give the correct answer
in the semiclassical limit and in the forward glory scattering \cite{LAM}.
As no probabilities can be assigned  to the individual terms
in Eq.(\ref{9.1}) we expect the proposed
estimate (\ref{9.2})  to be an improper average of some kind. 
Using the analogy with Eq.(\ref{7.11}) of the previous Section
we can rewrite Eq.(\ref{9.2}) as
\begin{equation}\label{9.3}
LAM(\theta) = \sum_{L=...-2,0,2...}Lw_L(\theta)
\end{equation}
where the normalised distribution $w_L(\theta)$ is given by 
\begin{equation}\label{9.4}
  w_L(\theta) =  Re \{ \Phi_L(\theta)/\sum_{L'=...-2,0,2...}\Phi_{L'}(\theta)\}
, \quad L= -2,0,2...
\end{equation}
and
\begin{equation}\label{9.5}
\Phi_L(\theta) \equiv \pi^{-1} \exp(iL\theta)\int_0^{\pi}
f(\theta ')\exp(-iL\theta')d\theta'.
\end{equation}
Note that the newly introduced quantity with the units
of angular momentum $L$ takes even integer values and 
is not identical to the total angular momentum $J$. As there are no {\it apriori}
restrictions on the phase of $ \Phi_L(\theta)$, $w_L(\theta)$ may
change sigh and the $LAM(\theta)$ is not, in general, 
required to take value within the range of the partial waves which
contribute to the scattering amplitude.
\newline A detailed discussion of the $LAM$ and its application to the analysis
of angular scattering will be given elsewhere.
As an illustration, we show in Fig. 3 $LAM(\theta)$ for the 
($v$,$j$ and $K$ are the vibrational, rotational and
helicity quantum numbers, respectively)
\begin{equation}\label{trans}
 v'=2,\quad j'=0, \quad K'=0 \leftarrow v=0,\quad j=0, \quad K=0
 \end{equation}
transition for the $F+H_2\rightarrow FH+H$ reaction \cite{FH2}
at the collision energy $E\approx 38 meV$.
Figure 3a shows the differential cross-section $\sigma(\theta) \equiv
|f(\theta)|^2|$ obtained by summing over $12$ partial waves, $0\le J\le 12$,
while $LAM(\theta)$ is plotted in Fig.3b.
and Fig. 3 c shows the distribution $w_L(\theta)$ in Eq.(\ref{9.4})
near the minimum of the DCS at  $\theta \approx 50^o$.
 
\section{Feynman amplitudes, the ABL rule and the three box case}
The last example given in this Section, although not directly related to weak
measurements or values, fits well within the general context of this 
paper.
Consider a three-slit experiment, in which an electron or a
photon may reach a detector through
slits $1$, $2$ and $3$. Let the amplitudes for the three
pathways be 
\begin{equation} \label{6.1}
A(1)=A(2)=1 \quad and \quad A(3)=-1,
\end{equation}
respectively. It is obvious now that since the contributions
from the routes $2$ and $3$ cancel each other, one may plug
the two slits without affecting the detector count.
It would be wrong, however, to conclude that the particle
always travels the route $1$, as the count would not be 
affected if the slits $1$ and $3$ were close instead.
In fact, this is the situation discussed in Section 2 (c).
The amplitude distribution for the slits can be written
as
\begin{equation}\label{6.2}
\Phi(f)= \delta(f-1)+\delta(f-2)-\delta(f-3),
\end{equation}
and the integral $\int \Phi(f) df$ gives the probability
amplitude to arrive at the detector. The distribution
alternates, we cannot
decide in a unique manner which two parts of the integral 
cancel each other, and must only conclude that it is not possible 
to determine through which slit the particle actully went.
The same gedankenexperiment can presented in a slightly
more intriguing form. Suppose that an meter determines
whether an electron goes through the slit $1$ but does not 
distinguish between the two other slits. Then each 
electron arriving at the screen will also be registered
at the slit $1$. Similarly, if the slit two 
is watched, the electron will always be found passing through
it.
All this is easily explained in terms of interfering and
exclusive alternatives (see Chapt.1, Sect.3 of  Ref. \cite{Feyn}). By watching the the slit
one we produce two alternative routes the
the detector: one (I) is through slit $1$ itself, and the other (II)
through both the slits $2$ and $3$, which remain interfering
alternatives and cannot be distinguished. Now we can use Feynman's prescription for assigning
probabilities: all interfering amplitudes must be added
coherently, and then the moduli of the sums must be squared,
\begin{equation} \label{6.3}
P(I)=|A(1)|^2/(|A(1)|^2+|A(2)+A(3)|^2)=1
\end{equation}
\begin{equation} \label{6.4}
P(II)=|A(2)+A(3)|^2/(|A(1)|^2+|A(2)+A(3)|^2)=0
\end{equation}
which explicitly show that the route (II) is not travelled
due to the destructive interference.
Even though an observation is conducted in such a 
way that it does not change the detector count,
it changes the situation and fails to provide
a clue as to what 'actually' happens to an unobserved
particle.
Note that if all three amplitude are chosen to be positive,
shutting any two slits would always effect
the detector count.
\newline
Consider further
\cite{3B1,AhBOOK} a three level system with a zero Hamiltonian
$\hat{H} \equiv 0$ is prepared and post-selected in the states
\begin{equation}\label{6.5}
|\Psi_0\ra = (3)^{-1/2}(|1\ra+|2\ra+|3\ra)
\end{equation}
and 
\begin{equation} \label{6.6}
|\Psi_1\ra = (3)^{-1/2}(|1\ra+|2\ra-|3\ra),
\end{equation}
respectively.
Between the preparation and the post-selection, the
projector on the first state, $\hat{P}_1 \equiv |1\ra \la 1|$
is accurately measured.
According to Eq.(\ref{3.6}), there
are only three paths connecting the initial
and final states $\Psi_0|\ra$ and $\Psi_1|\ra$,
\begin{equation} \label{6.8}
a_1(t')=1, \quad a_2(t')=2 \quad and \quad a_3(t')=3,
\end{equation}
with the corresponding amplitudes
given by
\begin{equation} \label{6.9}
A(n)= \la \Psi_1|n\ra \la n|\Psi_0\ra, \quad, n=1,2,3.
\end{equation}
The operator $\hat{P}_1$ has one simple
and one doubly degenerate eigenvalues
of $1$ and $0$, respectively, so that,
as was shown in Sect. 6 of Ref.\cite{SR1}, its
 measurement  destroys coherence between 
the paths in the same way as observing 
the particle passing through the first slit in the 
three-slit experiment. 
Thus, inserting Eq.(\ref{6.9}) into Eq.(\ref{6.3})
yields
\begin{equation} \label{6.10}
P(1)=|\la \Phi_1|1\ra \la 1|\Psi_0\ra|^2/(|\la \Psi_1|1\ra \la 1|\Psi_0\ra|^2 +
|\la \Psi_2|2\ra \la 1|\Psi_0\ra +\la \Psi_1|3\ra \la 3|\Psi_0\ra|^2)=1.
\end{equation}
Similarly, one always finds the particle in the second
'box' if the projector $\hat{P}_2 \equiv |2\ra \la 2|$
is measured instead, 
\begin{equation} \label{6.7}
P(2)\equiv 1/3/(1/3+1/3-1/3)=1
\end{equation}
Equation (\ref{6.10}) is the Aharonov, Leibowitz and Bergmann (ABL)
rule \cite{ABL1} for an operator with degenerate eigenvalues (see
Eq.(5) of Ref.\cite{3B1}).
Note that in our analysis the ABL rule is a simple
consequence of Feynman's prescription for adding probability 
amplitudes (see Sect. 1-7 of \cite{FeynL}) and does not rely a time symmetric formulation
of quantum mechanics employed in \cite{3B1}.
Also the Feynman's uncertainty principle suggests a different
interpretation of just described 'three box case'.
The authors of \cite{AhBOOK} note that
since the measurement of $\hat{P}_1$ and
$\hat{P}_2$, ('opening boxes 1 and 2' in the terminology
of \cite{AhBOOK})
always yield positive results, a
particle subjected to the boundary
conditions (\ref{6.5})-(\ref{6.6}) exists, at any 
intermediate time in two 'boxes' simultaneously.
Alternatively, it can be argued that the measurements
of the two projectors correspond to two distinct 
physical situations which, in turn, provide no 
clue as to where the particle actually is when
no measurement is conducted and all three 
pathways remain interfering alternatives.

\section{Conclusions and discussion}

In summary, quantum mechanics can be seen to operate by 
assigning probability amplitudes to scenarios or pathways
which can be interpreted as classical outcomes. 
Some of the scenarios are exclusive by nature, 
some are normally interfering but can be made exclusive
by coupling the system to a meter and some, it appears,
cannot be made exclusive at all, i.g., because a suitable
meter cannot be constructed. 
In general one  wishes then to know how many outcomes are there and
what is the likelihood of the realisation of a particular one.
A quantum measurement can be seen as performing this
taks by 
labelling the pathways by  some variable $f$ and
then analysing the moments of its distribution.
For exclusive scenarios, e.g., different values
of a variable $\hat{A}$ in the presence of an
accurate von Neumann meter,
 a proper probability 
distribution exists {\it apriori}.
However, some  phenomena such as
the interference pattern in a double
slit experiment or tunnelling transmission
across a potential barrier rely on constructive 
or destructive interference between the relevant pathways.
According to the Feynman's  uncertainty
principle, interfering scenarios cannot be told apart and
form, therefore, a single indivisible pathway connecting 
the initial and final states of a system.
Mathematically, the principle arises form the alternating
nature of the probability amplitudes responsible
for cancellation between the pathways, which, 
in turn, forbids the identification of the main contributor(s) 
to the transition.
Accordingly, we have observed, that an attempt to assign a mean value
to  $f$ when it labels  interfering alternatives,
be it by performing a weak von Neumann-like
measurement, by extending to the quantum context
a suitable classical procedure, as in the case of the
Wigner-Eisenbad phase time, or by postulating 
of an expression with an appropriate classical
limit, as in the case of the LAM, leads to an improper 
complex weak value $\bar{f}$  (\ref{0.1}).
This can, indeed, be expected, as in the absence of 
probabilities, $\bar{f}$ is the only average one can construct
from a real variable and complex probability amplitudes.
We note further that, contrary to what has been claimed by several authors
\cite{Rev1, Rev2},
the complexity of $\bar{f}$ is not in itself an obstacle 
to its interpretation, as the experiment always
dictates which part (s)  $\bar{f}$ (in our case, $Re\bar{f}$, for 
$Im\bar{f}$ and $|\bar{f}|$ see, for example \cite{SCx})
 should be used
to produce the required real answer.
\newline
A far more serious problem is that  $Re\bar{f}$
in an improper average obtained, in general, 
with an alternating distribution and has 
a number of undesirable properties  discussed
in Sect.2. In particular it may lie outside the 
region containing the support of the amplitude 
distribution, e.g.,  the spectrum of the measured
variable in the case of an impulsive von Neumann measurement
or the range of the time delays prescribed by the causality
in the case of the phase time,
and take anomalous
large valued of either sign even when this support is bounded.
\newline
Just because improper averages can take values which appear
unreasonable does not mean that they always do that.
In particular the amplitude distribution employed 
in their construction may or may not be improper
for all transitions, 
just as for some selected states the Wigner function $W(p,x)$ does
not always take negative values.
In general, quantum interference hides the information
about the range of the values contributing to the transition in a way that
 one can never 'trust' a weak value to represent the centroid of the range
without a detailed inspection of the distribution itself.
Of course, if such an inspection is possible, there is no longer
a need to evaluate the mean (\ref{0.1}).
\newline
In the end one cannot avoid asking of whether the weak values
should be treated as 'true' properties of a system 
in the presence of interference, or a manifestation
of a failure of a measurement designed to defy the uncertainty
principle.
Both points of view are, in principle, possible.
The former, expressed in \cite{Ah1,AhBOOK} is reinforced by the notion that a weak
value may be obtained in an act of measurement
and, therefore, provides
the only answer to the question about the value  taken by a variable 
in the presence of interference. There is  also no other
'correct' answer to refute it. 
However, a suggestion that with only 
two slits present an electron passes on average through the
slit number $-8$, or that a tunnelling particle spends on 
average a zero time within a barrier thereby defying the relativity,
clearly requires further clarification.
The explanation that the weak value 
is not actually tied to the range of the values contributing 
to the transition (numbers 1 and 2 of the slits or purely 
non-negative time delays in the case of the barrier)
seriously diminishes the value of the information a weak measurement
can provide.
\newline  
An alternative view can be summarised as follows.
 Interfering pathways 
cannot be told apart without destroying coherence between them and,
with it, the studied transition.
With the information destroyed by interference, a suitable answer 
to the above question simply does not exist. 
If one insists, e.g., by employing an extremely
inaccurate 'weak' meter, both the theory and experiment provide, much like
a politician or a manager, 
 a kind of non-answer, not necessarily related
to what has been asked.
\newline
In a similar manner Feynman's  uncertainty principle
can be used to 'resolve' the three box paradox of Sect. 7.
If no measurements are conducted, the particle
cannot be said to be in either particular box. 
Opening one of the box creates 
 a new physical situation and two
 exclusive pathways to which one can now assign
 probabilities which, however tell us nothing about
 the case when no measurements are made. 
The world 'resolve' is put in quotes because
the pathway analysis does not explain the 'logical difficulties' \cite{Feyn}
associated with quantum interference, but simply compacts them
into the Feynman's formulation of the uncertainty principle.
We conclude by quoting Feynman on the 
double-slit experiment:\cite{FeynL} :
"We choose to examine a phenomenon which
is impossible, {\it absolutely} impossible, to explain
in any classical way... In reality, it contains the
 {\it only} mystery.".

\newpage
\newpage
\begin{figure}[ht]
\epsfig{file=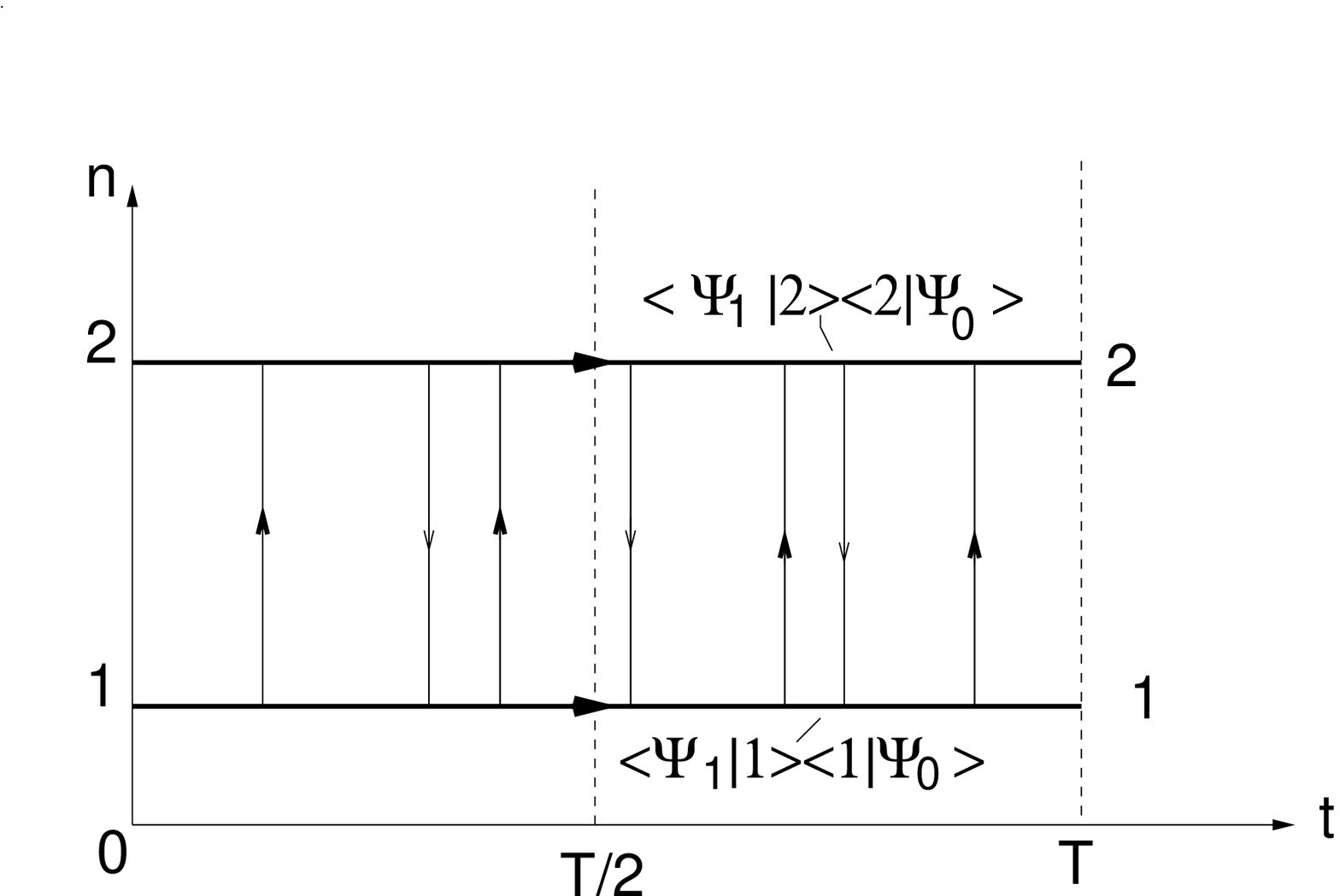, height=10cm, angle=0}
\vspace{0pt}
\caption{
\newline
Various virtual paths in Eq.(\ref{3.6}) contributing to the 
transition between $|\Psi_0\ra$ and $|\Psi_1\ra$ for  two-level
system.
For $\hat {H} =0$ only the two constant paths
(thick solid) have non-zero probability amplitudes.}
\end{figure}

\newpage
\newpage
\begin{figure}[ht]
\epsfig{file=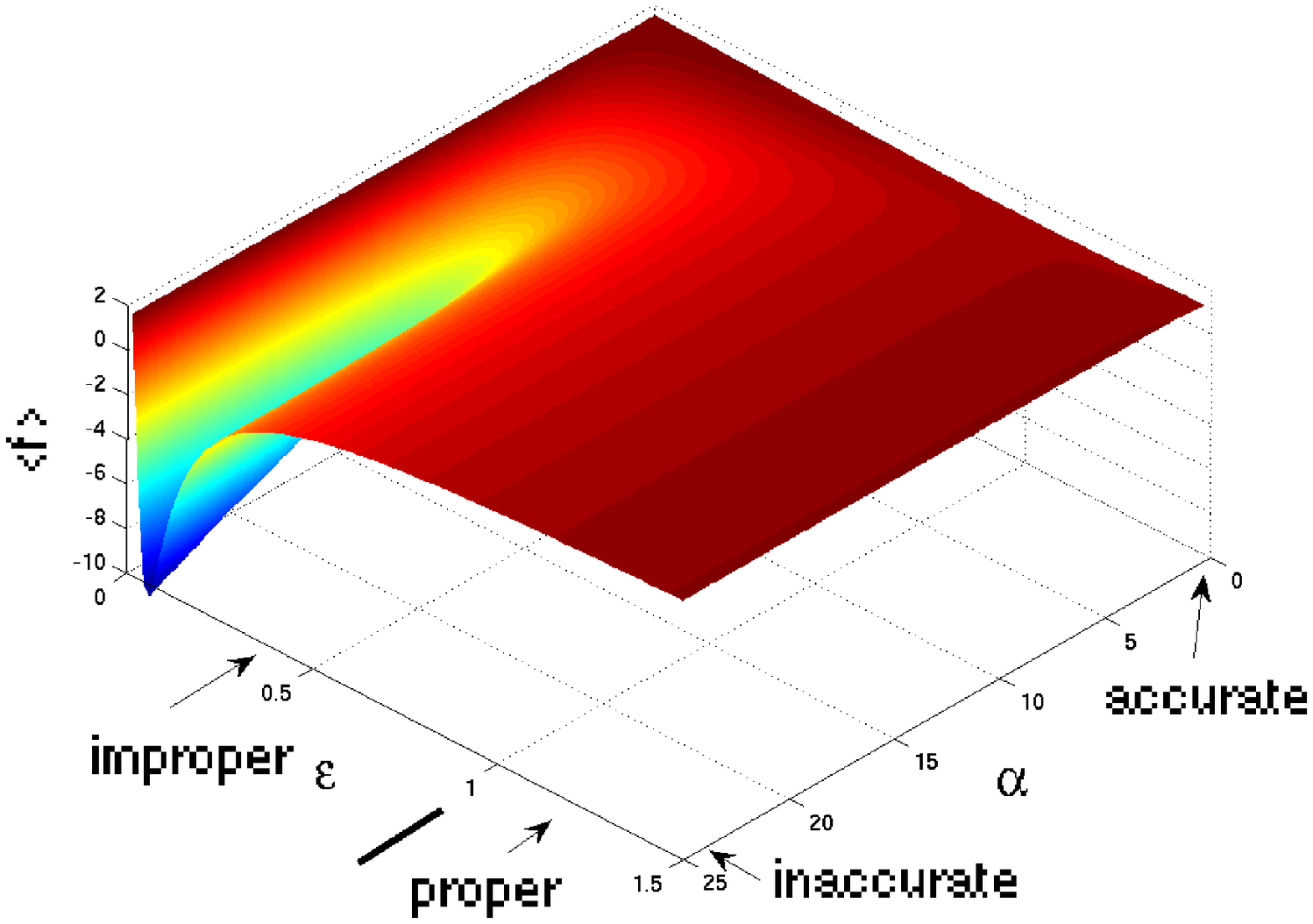, height=10cm, angle=0}
\vspace{0pt}
\caption{
\newline
The mean intermediate position of a two-level system in 
in the transition between the states  $|\Psi_0\ra$ and $|\Psi_1\ra$
in Eq.(\ref{5.2}) measured by a Gaussian meter 
with and accuracy $\alpha$} vs. $\epsilon$ and $\alpha$.
\end{figure}

\newpage
\newpage
\begin{figure}[ht]
\epsfig{file=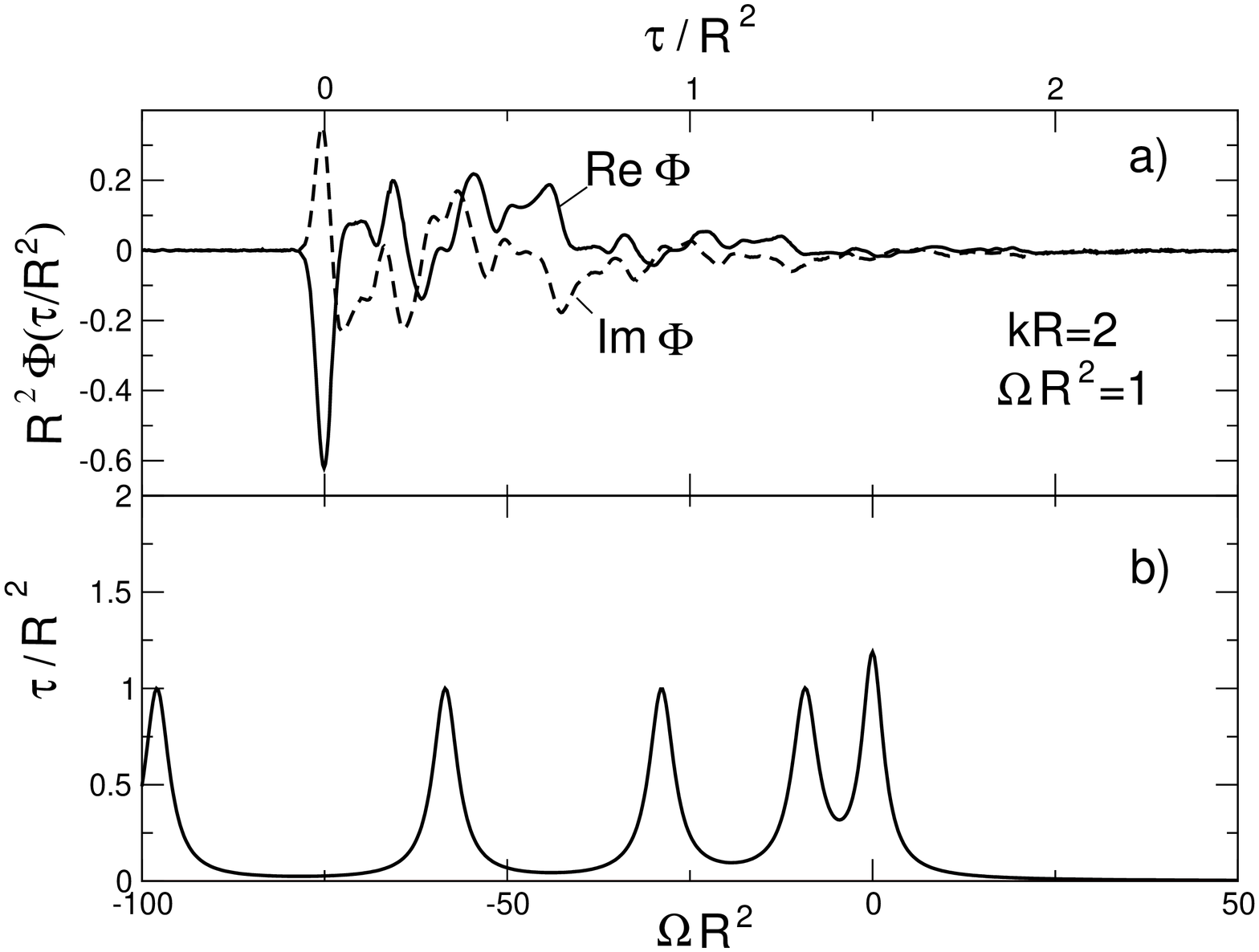, height=16cm, angle=-90}
\vspace{0pt}
\caption{
\newline
a) Real and imaginary parts of the collision 
(traversal) time distribution $\Phi(\tau)$ in Eq.(\ref{5.19})
(smeared with a Gaussian with the width
$\Delta (\tau)/R^2)=0.03$)
for a rectangular potential $V(r)$ of a height $\Omega$
and radius $R$.
\newline
b) The weak value $\bar{\tau}$ vs. $\Omega$ for the above potential.}
\end{figure}

\newpage
\newpage
\begin{figure}[ht]
\epsfig{file=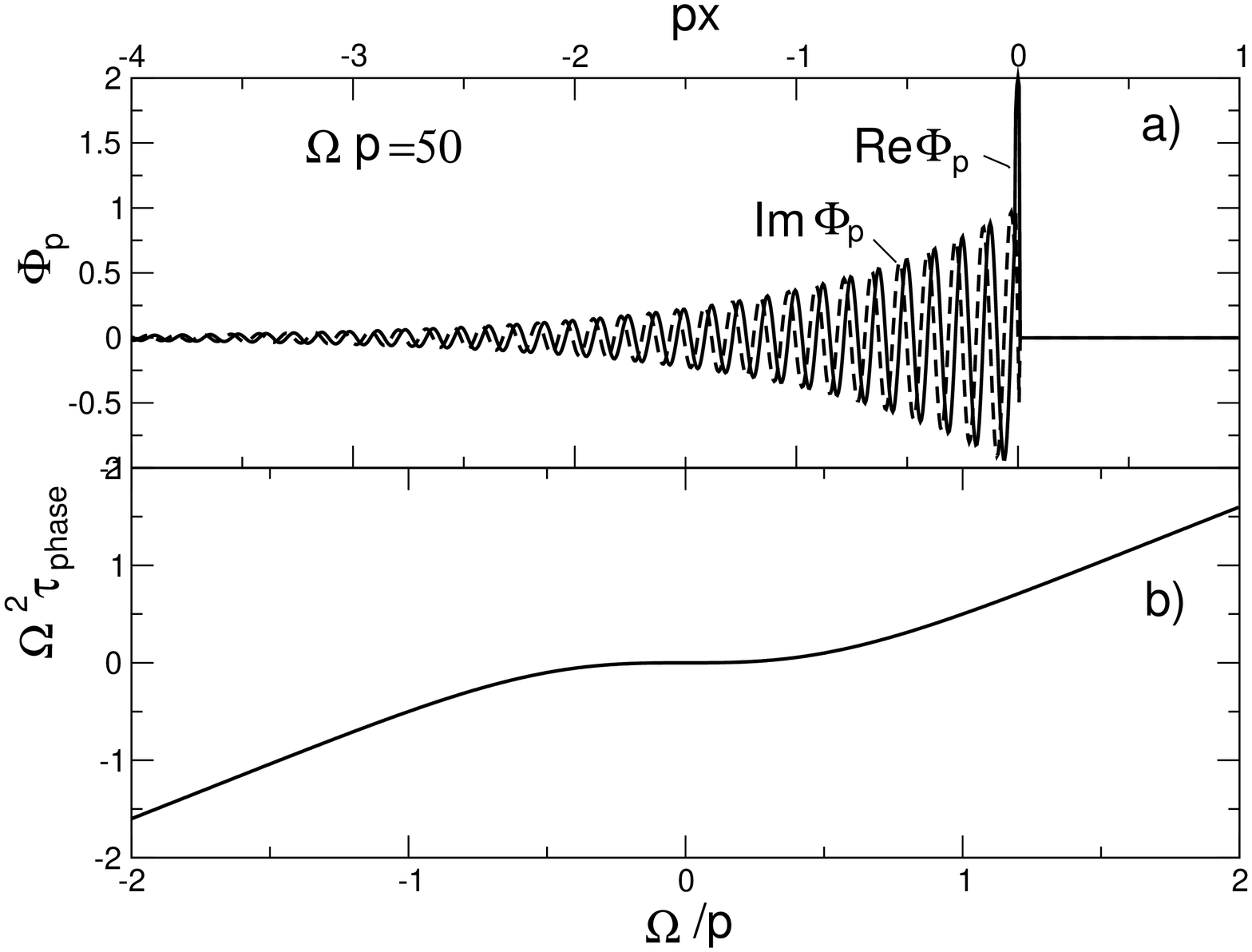, height=16cm, angle=-90}
\vspace{0pt}
\caption{
\newline
a) Real and imaginary parts of the amplitude distribution $\Phi(x)$ in Eq.(\ref{7.6})
for a thin barrier $V(x)=\Omega \delta (x)$.
\newline
b) The phase time $\tau_{phase}$ vs. $\Omega /p$}
\end{figure}

\newpage
\newpage
\newpage
\begin{figure}[ht]
\epsfig{file=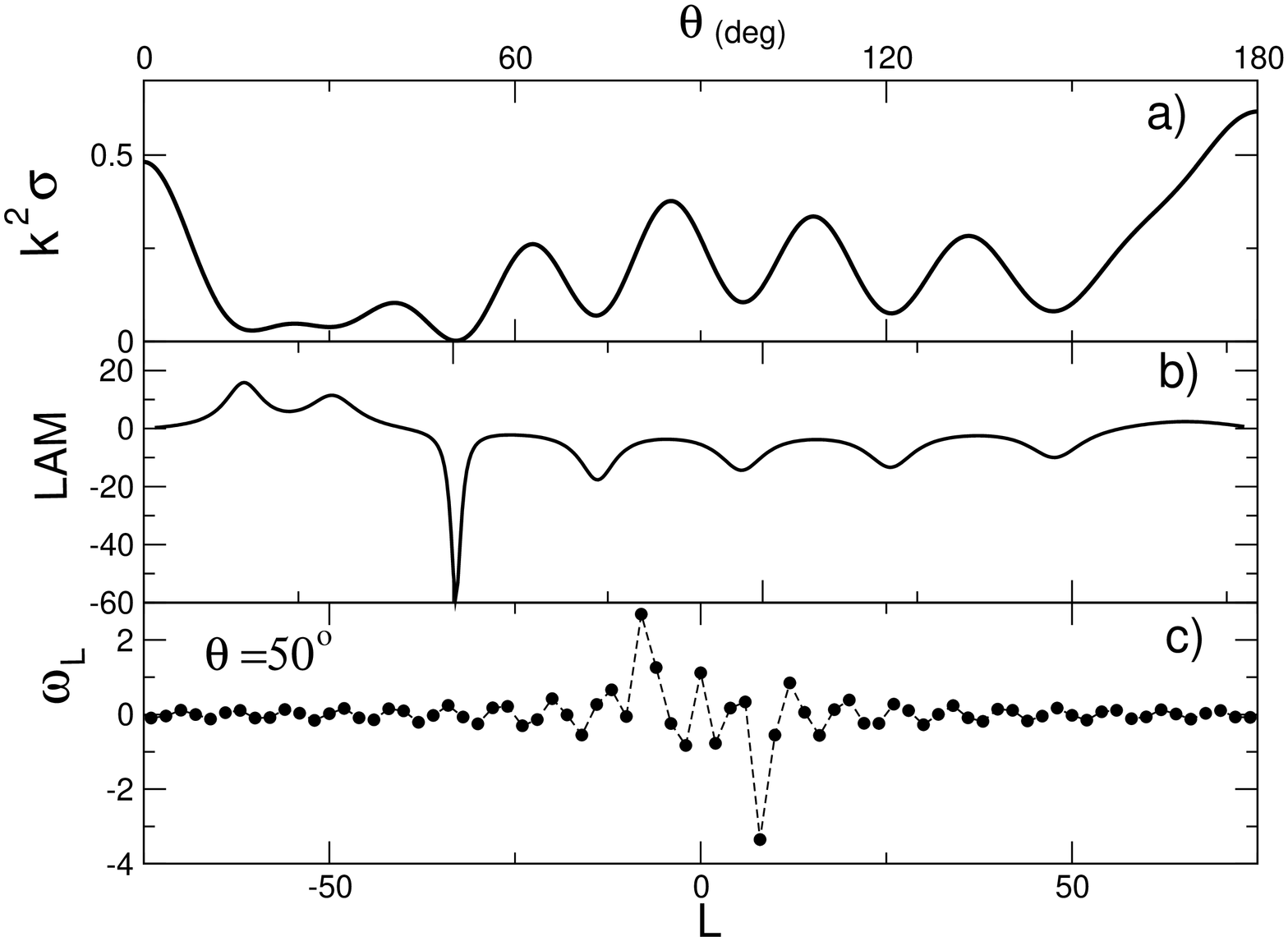, height=16cm, angle=-90}
\vspace{0pt}
\caption{
\newline
a) The $F+H_2$ reactive DCS for the  transition
(\ref{trans}) 
\newline
b)$ LAM(\theta) $ for the transition (\ref{trans})
\newline
c) The improper distribution $w_L(\theta)$ in Eq.(\ref{9.4}) vs. $L$
for $\theta \approx 50^o$}.
\end{figure}


\begin{thebibliography}{8.}
\bibitem{FOOT1}  A conventional von Neumann meter is a particular case, obtained if for 
 the choice of the switching function $\beta (t) \approx \delta(t)$ \cite{SR1}.
\bibitem{FOOT2}  Note that Eq.(\ref{5.18a})
results from choosing the operator $\hat{A}$ in Eq.(\ref{3.8}) 
to be the projector $\hat {P}_R$ onto the interior of the sphere,
and  $\beta(t')\equiv 1$. Since  $\hat {P}_R$ commutes with the position
$\vec{r}$, restricted sum (\ref{3.8}) can also be written as a sum over
Feynman paths (see Sect.VI of Ref.\cite{SR1}).
 \cite{SR1}.
\end{thebibliography}
\end{document}